\documentclass[10pt,prl,twocolumn,superscriptaddress,showpacs,letterpaper]{revtex4}

\usepackage{graphicx}
\usepackage{amssymb}
\usepackage{times}
\usepackage{amsmath}

\def\ket#1{\left\lvert {#1} \right\rangle}

\begin{document}

\title{Experimental Demonstration of Post-Selection based Continuous Variable Quantum Key Distribution in the Presence of Gaussian Noise}

\author{Thomas~Symul} \affiliation{Quantum Optics Group, Department
of Physics, Faculty of Science, Australian National University,
ACT 0200, Australia}

\author{Daniel~J.~Alton} \affiliation{Quantum Optics Group, Department
of Physics, Faculty of Science, Australian National University, ACT
0200, Australia}

\author{Syed~M.~Assad} \affiliation{Quantum Optics Group, Department
of Physics, Faculty of Science, Australian National University, ACT
0200, Australia}

\author{Andrew~M.~Lance} \affiliation{Quantum Optics Group, Department
of Physics, Faculty of Science, Australian National University, ACT
0200, Australia}

\author{Christian~Weedbrook} \affiliation{Quantum Optics Group, Department
of Physics, Faculty of Science, Australian National University,
ACT 0200, Australia} \affiliation{Department of Physics,
University of Queensland, St Lucia, Queensland 4072, Australia}

\author{Timothy~C.~Ralph} \affiliation{Department of Physics, University of Queensland, St Lucia, Queensland 4072, Australia}

\author{Ping~Koy~Lam} \affiliation{Quantum Optics Group, Department
of Physics, Faculty of Science, Australian National University,
ACT 0200, Australia}

\date{\today}

\begin{abstract}

In realistic continuous variable quantum key distribution protocols,
an eavesdropper may exploit the additional Gaussian noise generated
during transmission to mask her presence. We present a theoretical
framework for a post-selection based protocol which explicitly takes
into account excess Gaussian noise. We derive a quantitative
expression of the secret key rates based on the Levitin and Holevo
bounds. We experimentally demonstrate that the post-selection based
scheme is still secure against both individual and collective
Gaussian attacks in the presence of this excess noise.

\end{abstract}


\maketitle

Continuous variable quantum key distribution (CV-QKD) \cite{Bra03}
was introduced 
as an alternative to the original discrete variable single photon
schemes \cite{Gis02}. CV-QKD promises to offer higher secret key
rates, better detection efficiencies and higher bandwidths than its
single photon counterpart and is easily adapted to current
communication systems. Currently the two main protocols in CV-QKD
are post-selection (PS) \cite{Sil02} and reverse reconciliation (RR)
\cite{Gro03b}. These protocols are based on the random Gaussian
modulation of coherent states using either homodyne \cite{Gro03b} or
heterodyne \cite{Wee04} detection and both have been experimentally
demonstrated \cite{Gro03,Lod05,Lan05,Lod07,Lor06}. At present
PS-based CV-QKD has practical advantages in terms of key
distillation and has been demonstrated experimentally for up to
$90\%$ channel loss \cite{Lan05}.

Reverse reconciliation CV-QKD, due to its inherent nature, easily
incorporates excess noise into the protocols, and security proof
have been demonstrated in the case of individual Gaussian attacks
\cite{Gro03b, Wee04}, non-Gaussian attacks \cite{Gro04}, collective
attacks \cite{Gro05,Nav05} (with their Gaussian optimality
\cite{Nav06}) and coherent states using homodyne detection
\cite{Ibl04}. For PS CV-QKD, the addition of excess noise into the
analysis is quite difficult. The original protocol \cite{Sil02} only
considered pure or vacuum states in its scheme and so far all
post-selection protocols since have concentrated on the unrealistic
case of zero excess noise \cite{Nam04,Lan05,Nam06}. Recently
however, excess noise using a hybrid protocol, consisting of both
post-selection and either direct or reverse reconciliation, was
considered for the case of collective attacks \cite{Hei06}.

In this paper, we present a protocol for calculating the effect of
excess Gaussian noise (EGN) on post-selection where two way
classical communication is permitted, and show its security when
considering either individual or collective attacks. We apply our
analysis to an experimental demonstration and conclude that good key
rates can be obtained under the realistic condition of channel with
loss and excess Gaussian noise.

\begin{figure}[h]
  \begin{center}  
    \includegraphics[width=8.5cm]{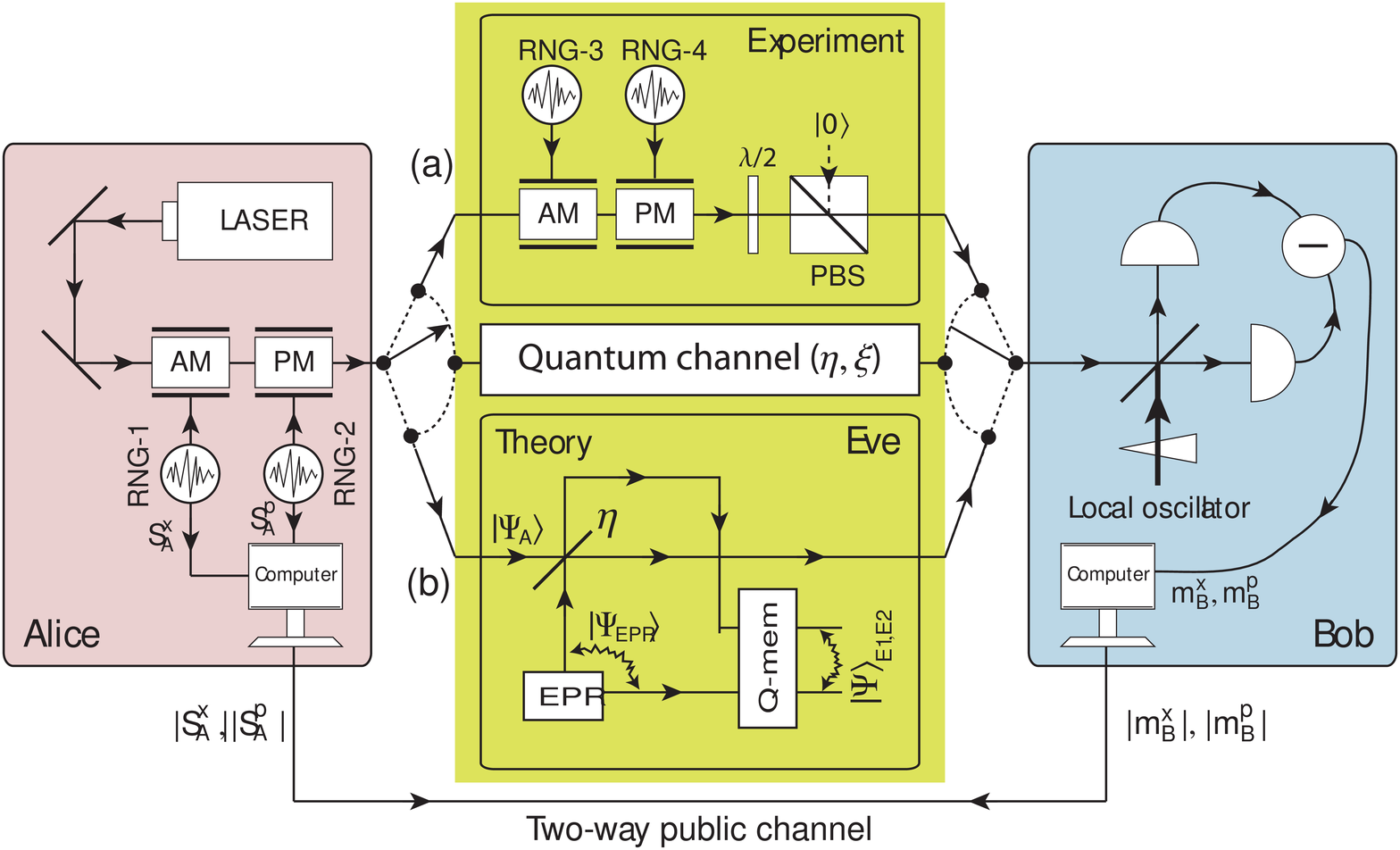}
  \end{center}
\caption{\small (color online). Schematic of setup. Quantum channel
of transmission $\eta$ and excess noise $\xi$ is simulated
experimentally (a) and analyzed theoretically for the entangling
cloner attack (b). $\lambda/2$: half waveplate; PBS: polarizing
beam-splitter; AM/PM: amplitude/phase modulators; RNG: independent
white noise generators; EPR: Entanglement source; Q-mem: quantum
memory.} \label{fig:Schematic}
\end{figure}

We extend the original PS CV-QKD protocol \cite{Sil02} as follows.
The sender, Alice draws two random numbers $S^x_A$ and $S^p_A$ from
Gaussian distributions of variances $V^x_A$ and $V^p_A$
respectively, which she encodes on the amplitude ($x$) and phase
($p$) of a coherent beam. Each encoding $(S^x_A,S^p_A)$ represent a
pair of bits whose value is fixed by the sign of the encoding. The
modulated Gaussian beam is then transmitted to the receiver, Bob,
through a lossy and noisy Gaussian channel with transmission $\eta$
and variance of EGN $\xi$. Bob receives a Gaussian mixed state
$\hat{\rho}_B$ with variance $V^{x,p}_B\!=\!\eta
V^{x,p}_A\!+\!1\!+\!\xi$, and then randomly measures either the
amplitude $m^x_B$ or phase $m^p_B$ quadratures of this mixed state.
As both amplitude and phase play the same role, we will only
explicitly consider one quadrature for the rest of this paper, and
denote Alice's encoding and Bob's measurement as $S_A$ and $m_B$
respectively. The probability that Bob measures a particular value
$m_B$ given that Alice used a particular encoding $S_A$ is given by
the conditional probability,
\begin{equation}\label{eqn:pmBSa}
p(m_B|S_A)=\frac{e^{- (m_B-\sqrt{\eta}S_A)^2/(2(1+\xi)V_V)}}{\sqrt{2
\pi (1+\xi)V_V}}
\end{equation}
where $V_V$ is the variance of the vacuum noise. Note that in this
paper the vacuum noise is normalized to $V_V=1$. The error rate in
Bob deciding whether Alice encoded positively or negatively is thus
given by
\begin{equation}
P_e=\frac{1}{1+e^{2\sqrt{\eta}\frac{\left|S_A
m_B\right|}{(1+\xi)V_V}}}
\end{equation}
The mutual information rate between Alice and Bob is given as a
function of this error probability using the Shannon formula
\cite{Sha48} $I_{AB}=\Phi(1-2P_e)$ where
\begin{equation} \Phi(x)=\frac{1}{2} \left[ (1+x)\log_2(1+x)+
(1-x)\log_2(1-x)\right]
\end{equation}
Bob then informs Alice over a public channel which quadrature he
measured and at what time interval. Alice and Bob then both announce
the absolute values of their encodings $|S_A|$ and measurement
results $|m_B|$ respectively. This is in contrast to previous zero
excess noise protocols where only Alice announces her absolute value
\cite{Sil02,Lan05}. Alice and Bob then post-select information for
which they have a mutual information advantage over Eve and discard
information for which they do not. Alice and Bob also choose a
random subset of data to characterize the channel loss $\eta$, the
EGN $\xi$ and check that the statistics are Gaussian. Finally Alice
and Bob use a two way reconciliation algorithm to reconcile their
data. 

As with any type of eavesdropping attack, we assume any EGN on the
quantum channel is always attributed to, and controlled by, the
eavesdropper, Eve. The fact that there exists excess noise on the
channel allows Eve to be entangled to Bob. There exists a known
upper bound $\xi < 2 \eta$ \cite{Nam04} to the amount of EGN $\xi$
that can be added on a channel of transmission $\eta$ above which
Alice and Bob's quantum correlation cease to exist \cite{Gro03b}. We
consider here the entangling cloner attack which has been shown
optimal for PS CVQKD with collective attacks in the presence of EGN
\cite{Hei06}. The entangling cloner attack (see
Fig.~\ref{fig:Schematic}~(b)) consists of Eve replacing the lossy
line by a beam splitter of transmission $\eta$ where one of the
inputs is Alice's initial state in a quadrature basis given by
\begin{equation}
\ket{\psi_A} = (2\pi)^{-{1/4}}\int dx_1
e^{-\frac{1}{4}(x_1-S_A)^2}\ket{x_1}
\end{equation}
and the second input is one arm of an entangled state Eve has
created given by
\begin{eqnarray}
 \ket{\psi_{\rm EPR}} = \!\!\!&&\!\!\! \frac {1}{\sqrt{2 \pi}} \int \!\!\!\int \!\!dx_2 dx_3 e^{\frac{1}{4}(-V_s x_2^{2}- x_3^2/V_s)}  \nonumber \\
 \!\!\!&&\!\!\! \ket{\frac{1}{\sqrt2} (x_2\! +\! x_3)} \ket{\frac{1}{\sqrt2} (x_2\!-\!x_3)}
\end{eqnarray}
where $1/2\left(V_s+1/V_s\right)=\left(1-\eta +\xi\right)/(1-\eta)$
is the variance of the entangled beam she injects to simulate the
EGN $\xi$. Eve keeps one of the entangled beams (denoted $E_1$) and
one of the outputs of the beam splitter (denoted $E_2$) while she
sends the remaining output to Bob (denoted $B$) through a perfect
noiseless and lossless line. When Bob performs his homodyne
measurement and announces its absolute value $|m_B|$, Eve's state
collapses to one of the four possible pure state given by $\left|
\psi^a_b \right>_{E_1,E_2}$, where the superscript $a\!=\!0,1$
refers to Alice's encoded bit and the subscript $b\!=\!0,1$ to Bob's
measured bit.
\begin{eqnarray}\label{eqn:psiE1E2}&& \left| \psi^a_b \right>_{E_1,E_2} = \frac{1}{\sqrt{\eta}(2 \pi)^\frac{3}{4}}\int\!\!\!\int dx_2 dx_3 \nonumber \\
\!\!\!&&\!\!\! \qquad
e^{-\frac{1}{4}\left[((-1)^b\frac{|m_B|}{\sqrt{\eta}}-(-1)^{a}
|S_A|-\sqrt{\frac{1-\eta}{2
\eta}}(x_3-x_2))^2+x^2_2 V_s+x^2_3/V_s\right]}\nonumber \\
\!\!\!&&\!\!\! \qquad \left|
-(-1)^{b}\sqrt{\frac{1-\eta}{\eta}}|m_B|-\sqrt{\frac{1}{2
\eta}}(x_2-x_3) \right>_{E_2}\nonumber \\
\!\!\!&&\!\!\! \qquad \left|\frac{1}{\sqrt{2}}(x_2+x_3)
\right>_{E_1}
\end{eqnarray}
Note that this state is not normalised,
$\left<\psi|\psi\right>=p_{m_B|S_A}$ given by Eq.~(\ref{eqn:pmBSa}).
The amount of secure bits that Alice and Bob can extract for each
transmission is given by
$\max\{0,I_{AB}\!-\!\max\{I_{AE},I_{BE}\}\}$. Eve chooses to
maximise her information with either Alice or Bob depending on which
will give her the greater information. If Eve decides to attack
Alice, she needs to distinguish between the states
$\rho_{AE}^a\!=\!\left|\psi^a_0\right>\left<\psi^a_0\right|\!+\!\left|\psi^a_1\right>\left<\psi^a_1\right|$.
To attack Bob, she needs to distinguish between the states
$\rho_{BE}^b\!=\!\left|\psi^0_b\right>\left<\psi^0_b\right|\!+\!\left|\psi^1_b\right>\left<\psi^1_b\right|$.

The inner products between these states can be computed explicitly
by performing the Gaussian integrations in Eq.~(\ref{eqn:psiE1E2}).
For example, the four terms that distinguishes Eve's input for
attacking Alice from her inputs for attacking Bob are:
\begin{eqnarray}
\left<\psi^0_0|\psi^1_0 \right> = \left<\psi^1_1|\psi^0_1 \right>
&=&\frac{{\exp}\left[-\frac{ m_B^2 +
  (1+\xi) S_A^2}{2(1+\xi)} \right]}{\sqrt{2\pi(1+\xi)}}
  \\
 \left<\psi^0_0|\psi^0_1 \right>\!
=\left<\psi^1_1|\psi^1_0 \right>&=&
  \frac{{\exp}\left[-\frac{(1+\xi)^2 m_B^2 +
  \eta S_A^2}{2(1+\xi)}
  \right]}{\sqrt{2\pi(1+\xi)}}
\end{eqnarray}
We see that at the critical value of $m^c_B =
\sqrt{\frac{1+\xi-\eta}{(1+\xi)^2-1}} S_A$, all the above inner
products are equal. Eve's input state for attacking Alice is
unitarily equivalent to that for attacking Bob, and hence her
accessible information with Alice is exactly the same as with Bob:
$I_{AE} =I_{BE}$. When $m_B > m^c_B$, Eve would gain more
information by attacking Bob while below this line she stands to
gain more by attacking Alice.

Given Eve's two input states, we need to find her accessible
information. If this is smaller than $I_{AB}$, Alice and Bob keep
the bit and distill a key from it. Our task now is to find Eve's
accessible information for such states. We bound this information
from above for both individual and collective attacks.

A bound on Eve's accessible information $I_E^{(i)}$ in the case of
individual attacks is calculated by providing her with the knowledge
on whether Alice and Bob's bit values match or not. With this
information, Eve's input is reduced to two pure states. Her
accessible information is bounded by
\begin{equation} I_E^{(i)} = p_1 \Phi\left(\sqrt{1-f_1^2 }\right) +
p_2 \Phi\left(\sqrt{1-f_2^2} \right)\end{equation} where $p_1$ is
the probability that Alice and Bob obtains the same bits and $p_2$
is the probability that their bits differ, and
\begin{equation}
f_1=\frac{\left<\psi^0_0|\psi^1_1\right>}{\left<\psi^0_0|\psi^0_0\right>}\mbox{
and}
f_2=\frac{\left<\psi^0_1|\psi^1_0\right>}{\left<\psi^0_1|\psi^0_1\right>}
{\text ,}
\end{equation}
are the normalised inner products between the states that Eve
distinguishes \cite{Lev95}. We note that this bound corresponds to
the Levitin bound as given in \cite{Sil02} for the case of no added
noise.

\begin{figure}[h]
  \begin{center}  
    \includegraphics[width=8.5cm]{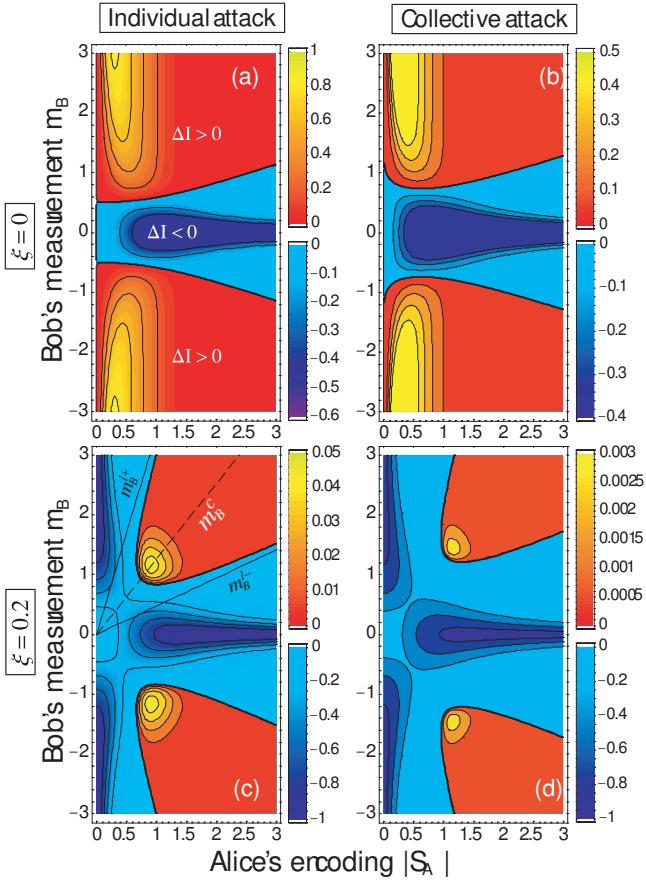}
  \end{center}
\caption{\small\small (color online). Post-selection regions at
$\eta\!=\!0.5$ are shown in red. Figures (a) and (b) show the
information rates $\Delta I\!=\!I_{AB}\!-\!I_E$ with no excess noise
for individual and collective attacks. Figures (c) and (d) is when
$\xi\!=\!0.2$. On the dashed line $m_B^c$ in figure (c), Eve can
obtain the same amount of information from Alice as she can from
Bob. The post-selection region asymptotes to the lines $m_B^{l
\pm}$.} \label{fig:ps}
\end{figure}

We apply Holevo's theorem \cite{Hol73} on Eve's input states,
$\rho_E$, to bound Eve's information in terms of the von Neumann
entropy, $S(\rho)$, and obtain the amount of information $I_E^{(c)}$
accessible by Eve when performin collective attacks
\begin{equation} I_E^{(c)} = S(\rho_E^0 +
\rho_E^1)- S(2 \rho_E^0)/2 -S(2 \rho_E^1)/2
\end{equation}
Figure~(\ref{fig:ps}) shows the difference in mutual information
from Bob's point of view when Alice announces $S_A$ for a fixed
value of $\eta$ and $\xi$. For each $\eta$ and $\xi$, Alice then
chooses the value of $V_A^{opt}$ such that the weighted integral
over the positive information region $\Omega$ given below is
maximised.
\begin{align}
\Delta I^{(i,c)} = \int_\Omega p(S_A,m_B) (I_{AB} -
I_{E}^{(i,c)})dm_B dS_A
\end{align}

In principle, as long as the post-selection region is non-empty,
Alice and Bob can always distill a finite amount of key. At a
certain noise threshold however, we expect that there will be no
more post-selectable region. This is clear for $\xi\!=\!2~\eta$
\cite{Nam04}, since then the state between Alice and Bob becomes
separable. In this case, Eve can do an intercept and resend attack
in which $I_E>I_{AB}$ for all values of $S_A$ and $m_B$.

But even before the separability limit is reached, the
post-selectable region may become empty. To analyse this, we
consider the case when $S_A$ is large. In such a case, Alice and Bob
would share the same bits with a high probability. Eve's accessible
information then tends to $\Phi \left( \sqrt{1-f_1^2} \right)$. In
this limit, Eve's input becomes ever closer to being just two
classical pure states and so Holevo's bound would tend to the same
limiting information. Equating this with $I_{AB}$, we obtain two
solutions for $m_B^{l\pm}$:
\begin{equation}
m_B^{l\pm}\!\!=\!\!\frac{\sqrt{\eta}(1\!+\!\xi)\!\!\pm\!\!\sqrt{\eta
(1\!+\!\xi)^2\! -\!
\xi(\xi\!+\!2)(\xi\!+\!1\!-\!\eta)}}{\xi(2\!+\!\xi)} S_A.
\end{equation}
In other words, the region of post-selectibility asymptotes to these
two lines as $S_A$ increases (see Fig.~\ref{fig:ps}). The noise
threshold $\xi_0$ over which the quantum channel is insecure is
obtained when the two lines $m_B^{l+}$ and $m_B^{l-}$ coincide such
that there is no more region of post-selectibility. This occurs when
$\eta(1+\xi_0)^2=\xi_0(\xi_0+2)(\xi_0+1-\eta)$.

\begin{figure}[h]
  \begin{center}  
    \includegraphics[width=8.5cm]{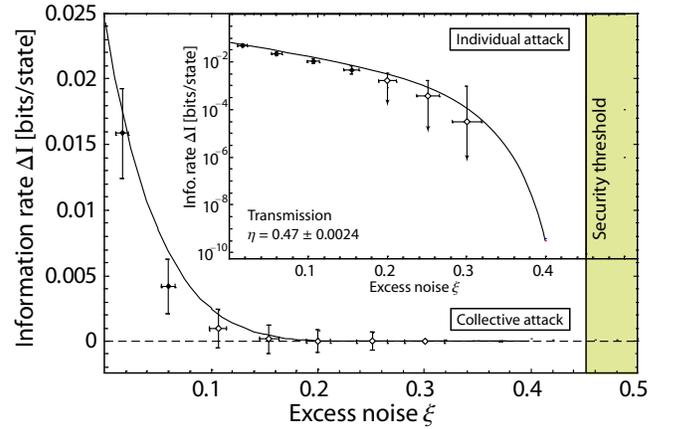}
  \end{center}
\caption{\small\small (color online). Experimental results
superimposed on theoretical lower bounds of secure key rates at
transmission $\eta=0.47\pm0.002$ when Eve does a collective attack
and an individual attack (inset). Unfilled data points with arrows
have error bars going to negative $\Delta I$. The shaded region
indicates the noise threshold for secure keys. The experimental
results were obtained using an encoding variance optimized for the
individual attack bound.} \label{fig:ExpLinLog}
\end{figure}

Fig.~\ref{fig:Schematic}(a) shows the schematic of our experiment.
In this setup we encoded keys on the amplitude quadrature and
simulated the loss of the line by using a variable attenuator and
the added noise by adding a random Gaussian signal onto the
amplitude quadrature on Alice's amplitude modulator. The transmitted
light is detected using a homodyne detection setup at Bob's station.
The two sets of time series, Alice's encoding $S_A$ and Bob's
measurement $m_B$ were analysed using the tools developed in
\cite{Lan05}. We note that extraction of the final key can be
performed using the methods described in \cite{Lan05} with an
average efficiency of 2 to 4 \% for all datasets with positive raw
information rates $\Delta I$.

Figure~\ref{fig:ExpLinLog} shows experimental results superimposed
onto theoretical bounds of total post-selected information rates
$\Delta I = I_{AB}-I_E$ at channel transmission $\eta = 47\%$ for
individual and collective attacks, as a function of channel EGN
$\xi$. The experimental mutual information rate between Alice and
Bob $I_{AB}$ is calculated by comparing the two signal-processed
time series $S^x_A$ and $m^x_B$. This quantity is less than the
theoretical calculation due to experimental imperfections associated
with the encoding (e.g. non-optimum encoding variance), detection
(e.g. homodyne inefficiency) and signal processing. Experimental
uncertainty is calculated for $I_{AB}$ due to the finite number of
data points. The information rate for Eve $I_E$ is calculated
theoretically, with error bars in $I_E$ calculated using the
uncertainties in channel transmission, EGN and Alice's variance
$V_A$.

The experimental data points $\Delta I_{\rm exp}$ are in good
agreement with the theoretical results. For some of the higher EGN
cases, the error bars extend towards the negative region. It should
be emphasized however, that this is mainly due to the finite number
of collected data that results in statistical uncertainties. In our
experiment, 2.4MBits of data were taken per run. The theoretical
curves for $\Delta I$ in Fig.~\ref{fig:ExpLinLog} monotonically
decreases until they reach exactly zero at the security threshold
line. No secure keys can be generated in the shaded region.

\begin{figure}[h]
  \begin{center}  
    \includegraphics[width=8.6cm]{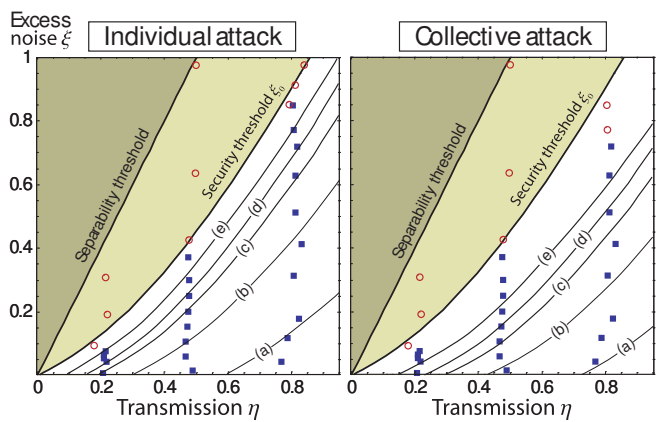}
  \end{center}
\caption{\small\small (color online). Experimental results
superimposed on theoretical contour plots of information rate after
post-selection ($\Delta$I) as a function of channel transmission
$\eta$ and EGN $\xi$. (a),(b),(c),(d),(e) indicates $\Delta
I=10^{-1},10^{-2},10^{-3},10^{-4},10^{-7}$. Filled (unfilled) data
points indicate $\Delta I_{\rm exp}>0$ ($\Delta I_{\rm exp}\leq 0$).
No secure keys can be generated in the shaded regions. Dark shade
indicates separability between Alice's and Bob's states.}
\label{fig:GeneralResults}
\end{figure}

Fig.~\ref{fig:GeneralResults} shows the experimental results
superimposed on contour plots of $\Delta I$ as a function of $\eta$
and $\xi$. Three sets of experimental runs were taken for
$\eta\!\approx0.2,0.5,0.8$. Filled and unfilled data points indicate
$\Delta I_{\rm exp}\!>\!0$ and $\Delta I_{\rm exp}\!\leq\!0$
respectively. We obtained positive information rates for
$\eta\!=\!0.2$ for $\xi\!=\!0.1$. In principle, lower $\eta$ is
attainable; the experimental demonstration for such cases is left
for future work.

In conclusion, we have extended the original post-selection protocol
\cite{Sil02} to take into account the effect of channel EGN for
individual and collective Gaussian attacks by an eavesdropper. In
both cases, we find that the scheme is still secure. We have also
presented an experimental demonstration, which verifies for the
first time that continuous variable quantum cryptography using
post-selection is secure in the presence of channel loss as well as
EGN. This is important since realistic laser sources and optical
fibers \cite{Lod05} inevitably inherit EGN. Reanalyzing our results
from \cite{Lan05} using the theory presented in this paper we
conclude that the small amount of EGN present in that experiment
would have had negligible effect on the key rates if properly
accounted for.

We thank Ch.~Silberhorn and N.~Lutkenhaus for useful discussions and
acknowledge financial support from the Australian Research Council
and the Department of Defence.

\end{document}